\documentclass[prl,graphicx,amsmath,amssymb,preprint]{revtex4-1}

\usepackage{graphicx}
\usepackage{color}

\draft 

\begin{document}


\title{Tachyons in ``momentum-space'' representation} 



\author{V. Aldaya}
\email{valdaya@iaa.es}
\affiliation{%
Instituto de Astrof\'\i sica de Andaluc\'\i a (IAA-CSIC), Glorieta de la Astronom\'\i a, E-18080 Granada, Spain
}%
\author{J. Guerrero}%
 \email{julio.guerrero@ujaen.es}
\affiliation{ 
Departamento de Matem\'aticas, Universidad de Ja\'en, Campus las Lagunillas, 23071 Ja\'en, Spain
}%
\altaffiliation[Also at ]{Institute Carlos I of Theoretical and Computational Physics, University of Granada, Fuentenueva s/n, 18071 Granada, Spain}

\author{F.F. L\'opez-Ruiz}
\email{paco.lopezruiz@uca.es}
 \affiliation{Departamento de F\'\i sica Aplicada, Universidad de C\'adiz, Campus de Puerto Real, E-11510 Puerto Real, C\'adiz, Spain}


\date{\today}

\begin{abstract}
The momentum space associated with tachyonic ``particles'' proves to be rather intricate, departing very much from the ordinary dual to Minkowski space directly parametrized by space-time translations of the Poincar\'e group.  In fact, although described by the constants of motion (Noether invariants) associated with space-time translations, they depend non-trivially on the parameters of the rotation subgroup. However, once the momentum space is parametrized by the Noether invariants, it behaves exactly as that of ordinary particles. On the other hand, the evolution parameter is no longer the one associated with time translation, whose Noether invariant, $P_o$, is now a basic one. Evolution takes place in a spatial direction. These facts not only make difficult the computation of the corresponding representation, but also force us to a sound revision of several traditional ingredients related to Cauchy hypersurface, scalar product and, of course, causality. After that, the theory becomes consistent and could shed new light on some special physical situations like inflation or traveling inside a black hole.

\end{abstract}

\pacs{Valid PACS appear here}

\maketitle 

\section{Introduction}
The tachyonic representations of the proper, orthochronous Poincar\'e group, $\mathcal{P}^\uparrow_+$, that is, representations associated with the one-sheeted hyperboloid as coadjoint orbit, were firstly considered by Wigner in 1939 \cite{Wigner}. The corresponding Quantum Theory was discarded from the beginning as being considered unphysical due mainly to the problems of causality, which arise when analyzed on the same footing as the other representations of $\mathcal P^\uparrow_+$. 

However, in recent days, many papers have been published where tachyons do appear although more usually involved in virtual scenarios \cite{Sen,Armoni,ellis,Landulfo,Dragan1,Dragan2,Paddy}. 

In this paper, we revisit the quantization of tachyons in quite a different scheme entirely based on symmetry grounds. We resort to a Group Approach to Quantization, which has been largely applied to many physical systems and has reported certain new light in more controversial situations (see, for instance, Refs.~\citep{oscilata,S3config,S3mom}), and recovered well-known results in more standard problems (see Refs.~\cite{2000,2004}). Here we accomplish in particular the quantization of tachyons in the momentum-space representation, poorly studied in the literature. The main advantage of the present approach is that the minimal, natural, independent arguments of the reduced wave functions depend on the Noether invariant associated with the time translations. This means that the time coordinate is not the evolution parameter and that the corresponding operator ($\hat{p}_o$) is not a function of any other operator (i.e., it is a basic operator); that is, wave functions are initially fixed by arbitrary functions on the Noether invariants (momenta) associated with time and two more $x$-parameters, and evolve on the third $x$-coordinate, but in a way that preserves the symmetry. We arrive, this way, at a quite different kinematics which entails quite different reasoning as far as velocity and/or causality is concerned. 

The paper is organized as follows. In Sec. 2 we propose the particular choice of the Poincar\'e group (central-)extension by $U(1)$ corresponding to the tachyonic representation, among other possibilities, and proceed with the computation of the wave functions in momentum space, as well as the expression of the quantum operators. In Sec. 3 we account for the way in which we approach the apparently more familiar realization on configuration space without entering explicit computations. Finally, Sec. 4 devotes some words to relevant physical considerations derived from the new results. An Appendix contains a very brief presentation of our general quantization scheme.  

\section{The momentum space representation} 
Let us write the Poincar\'e algebra exhibiting all possible central extensions that corresponds to the three classes of coadjoint orbits, that is, the two-sheeted hyperboloid, related to the representations associated with ordinary particles; the cone, related to photons; and the one-sheeted hyperboloid, corresponding to tachyonic elementary systems. We have:
\begin{align*}
	[\tilde X_{a^o},\tilde X_{a^i}] &=0, &[\tilde X_{a^i},\tilde X_{u^j}] &=-\delta_{ij} \tilde X_{a^o} - \lambda_o \Xi, \\ 
	[\tilde X_{a^o},\tilde X_{\epsilon^i}] &=0, 
	&[\tilde X_{a^o},\tilde X_{u^i}] &=-\tilde X_{a^i} - \lambda_i \Xi\\
	[\tilde X_{a^i},\tilde X_{a^j}] &=0, 
	&[\tilde X_{a^i},\tilde X_{\epsilon^j}] &=-\eta_{ij\cdot}^{\phantom{ij}k} \tilde X_{a^k} - \eta_{ij\cdot}^{\phantom{ij}k} \lambda_k \Xi \\
	[\tilde X_{\epsilon^i},\tilde X_{u^j}] &=\eta_{ij\cdot}^{\phantom{ij}k} \tilde X_{u^k}, 
	&[\tilde X_{\epsilon^i},\tilde X_{\epsilon^j}] &=\eta_{ij\cdot}^{\phantom{ij}k} \tilde X_{\epsilon^k} + \eta_{ij\cdot}^{\phantom{ij}k} n_k \Xi\\
	& &[\tilde X_{u^i},\tilde X_{u^j}] &=-\eta_{ij\cdot}^{\phantom{ij}k} \tilde X_{\epsilon^k} - \eta_{ij\cdot}^{\phantom{ij}k} n_k \Xi,
\end{align*}
where the parameters $(a^o,a^i)$ refer to space-time translation, $u^i$ to boosts and $\epsilon^i$ to rotations. $\Xi\equiv \tilde X_\zeta$ is the central generator, $\zeta \in U(1)$. It must be remarked that from now on  $\vec{u}\equiv c \sqrt{1+\frac{\vec{\alpha}^2}{4}}\vec{\alpha}$ is just the boost parameter (see Ref. \cite{AnnPhys}) and that quantities like $\vec p \equiv m \vec u$ and $p_o \equiv m u \sqrt{1+\frac{\vec{\alpha}^2}{2}}$ should not be identified with the four momentum of a particle. The proper constraint among momenta, satisfying the mass-shell condition will be established with the Noether invariants associated with the generators of space-time translations. 

The four-vector $(\lambda_o,\vec \lambda)$ defines a cocycle (see below) tied to the subgroup of the space-time translations and is associated with a class of representations according to its Lorentz character. Then, $(\lambda_o,0)$ corresponds to a representative of the ordinary mass (two-sheeted hyperboloid) class, $(\pm|\vec\lambda|,\vec \lambda)$ to photons and $(0,\vec\lambda)$ to a natural representative class of tachyons. The vector $\vec n$ defines a cocycle associated with the subgroup of rotations and is responsible for the spin of ordinary particles, the helicity of photons and something more involved for tachyons \cite{Moses}. Here we shall disregard non-trivial $\vec n$'s. Then, we proceed with the simplest choice corresponding to tachyons $(0,\vec\lambda)$ and $\vec n=0$. For the group law of the central extension $\tilde{\mathcal P}^\uparrow_+$of the Poincar\'e group $\mathcal P^\uparrow_+$, we adopt the standard group law like the one in Ref.~\cite{A26} added with the $U(1)$-group law: 
\begin{align}
	g''&=g' * g\,,   &g \in \mathcal P^\uparrow_+ \nonumber \\
	\zeta''&= \zeta' \zeta e^{i\frac{mc}{\hbar}\vec \lambda\cdot(\vec a''-\vec a'-\vec a)}\,,  &\zeta \in U(1)\,,
	\label{ley}
\end{align}
which corresponds to having chosen the cocycle (coboundary, indeed) generated by the function on $\mathcal P^\uparrow_+$: $\eta(g)=\frac{mc}{\hbar} \vec \lambda\cdot\vec a \Rightarrow \xi(g',g)= \eta(g'*g-g'-g)$, $g\in \mathcal P^\uparrow_+$. In the following, we shall choose units in which $\hbar=1$. 

In accordance with the cocycle in \eqref{ley},  we may derive the expression for left-invariant generators copying from Ref.~\cite{A26} all but the components in $\Xi$: 
\begin{align}
	\tilde X^L_{a^o}&= \frac{u^o}{c}\frac{\!\!\partial}{\partial a^o}+ \frac{\vec u}{c}\frac{\!\!\partial}{\partial \vec a}+ \vec \lambda\cdot\frac{\vec u}{c}\Xi \nonumber
\\
	\tilde X^L_{a^i}&= R(\vec \epsilon\,)^j_i\Big[\frac{\!\!\partial}{\partial a^j}+\frac{u_j}{c}\frac{\!\!\partial}{\partial a^o}+ 
	\frac{ u_j}{c(u_o+c)}\big(\vec u \cdot\frac{\!\!\partial}{\partial \vec a}\big)+
	(\mathbb I-R(\vec \epsilon\,)^{-1})_j^k \lambda_k \Xi+ \frac{(\vec \lambda \cdot \vec u)}{c(u_o+c)} u_j\Xi  \Big] \nonumber
	\\
	\tilde X^L_{u^i}&= R(\vec \epsilon\,)^j_i\Big[\frac{\!\!\partial}{\partial u^j}+\frac{u_j}{c(u_o+c)}\big(\vec u \cdot\frac{\!\!\partial}{\partial \vec u}\big)\Big]+ 
	\frac{1}{c(u_o+c)} \big[R(\vec \epsilon\,)^{-1} \vec u \wedge 
	X^{L\,SU(2)}_{\vec \epsilon}\big]_i
	\\
	\tilde X^L_{\vec \epsilon}&=  X^{L\,SU(2)}_{\vec \epsilon} \nonumber
	\\
	\tilde X^L_{\zeta}&= i\zeta \frac{\!\!\partial}{\partial \zeta}-i\zeta^*\frac{\!\!\partial}{\partial \zeta^*}\equiv \Xi\,, \nonumber
	\label{leftX}
\end{align}
where we have used the notation $\vec u$ for the boosts parameters instead of $\vec p \equiv m \vec u$ in order to avoid the use of the $m$, reminiscent of the mass of ordinary particles. We have also used the expression for the rotation matrices 
$R(\vec \epsilon\,)^i_j=(1-\frac{\vec{\epsilon}\,^2}{2})\delta^i_j-\sqrt{1-\frac{\vec{\epsilon}\,^2}{4}}\eta^i_{\cdot jk}+\frac{1}{2}\epsilon^i\epsilon_j$, where $\eta^i_{\cdot jk}$ stands for the antisymmetric  Levi-Civita (pseudo-)tensor. 

In the same way, the right-invariant generators follow: 
 \begin{align}
	\tilde X^R_{a^o}&= \frac{\!\!\partial}{\partial a^o} \nonumber
\\
	\tilde X^R_{\vec a}&= \frac{\!\!\partial}{\partial \vec a} \nonumber
\\
	\tilde X^R_{\vec u}&= \frac{u^o}{c}\frac{\!\!\partial}{\partial \vec u}-\frac{\vec u \wedge X^{R\,SU(2)}_{\vec \epsilon}}{c(u_o+c)}+\frac{a^o}{c}\frac{\!\!\partial}{\partial \vec a}+ 
	\frac{\vec a}{c}\frac{\!\!\partial}{\partial a^o} 
	+\frac{a^o}{c}\vec{\lambda}\;\Xi
\\
	\tilde X^R_{\vec \epsilon}&=  X^{R\,SU(2)}_{\vec \epsilon} 
	+ \vec a \wedge  \frac{\!\!\partial}{\partial \vec a}
	+ \vec u \wedge  \frac{\!\!\partial}{\partial \vec u}+
	\vec a \wedge \vec \lambda \Xi \nonumber
	\\
	\tilde X^R_{\zeta}&= i\zeta \frac{\!\!\partial}{\partial \zeta}-i\zeta^*\frac{\!\!\partial}{\partial \zeta^*}\equiv \Xi\,. \nonumber
	\label{rightX}
\end{align}
 
From \eqref{leftX}, the quantization form $\Theta \equiv \theta^{L(\zeta)}$ (the canonical left-invariant 1-form in the $\zeta$ direction) is derived:
\[
\Theta = - R(\vec \epsilon\,) \vec \lambda \cdot \frac{\vec u}{c} da^o - 
\Big[(\mathbb I-R(\vec \epsilon\,)) \vec \lambda - \frac{(R(\vec \epsilon\,)\vec \lambda \cdot \vec u)\vec u}{c(u_o+c)}\Big]\cdot d\vec a + \frac{d\zeta}{i\zeta}\,,
\]
and contracting it with the right-invariant generators, the Noether Invariants are found: 
\begin{align*}
	i_{\tilde X^R_{a^o}}\Theta &= -R(\vec \epsilon\,)\vec \lambda\cdot\frac{\vec u}{c}\equiv-P_o\,,\\
	i_{\tilde X^R_{\vec a}}\Theta &=  \big(R(\vec \epsilon\,)-\mathbb I\big)\vec \lambda + \frac{\big(R(\vec \epsilon\,)\vec \lambda\cdot\vec u\big)\vec u}{c(u^o+c)}\equiv\vec P-\vec \lambda\,,
	\quad
	\hbox{or}
	\quad
	i_{(\tilde X^R_{\vec a}+\vec \lambda\Xi)}\Theta =  R(\vec \epsilon\,)\vec \lambda + \frac{\big(R(\vec \epsilon\,)\vec \lambda\cdot\vec u\big)\vec u}{c(u^o+c)}\equiv\vec P\,,
	\quad
	\\
	i_{\tilde X^R_{\vec \epsilon}}\Theta &=  \vec a \wedge \Big[R(\vec \epsilon\,) \vec \lambda + \frac{\big(R(\vec \epsilon\,)\vec \lambda\cdot\vec u\big)\vec u}{c(u^o+c)}\Big]=\vec a\wedge\vec P\equiv\vec J\,,\\
	i_{\tilde X^R_{\vec u}}\Theta &= -\frac{1}{c^2}\big(R(\vec \epsilon\,)\vec \lambda\cdot\vec u\big)\vec a +  \Big[R(\vec \epsilon\,) \vec \lambda + \frac{\big(R(\vec \epsilon\,)\vec \lambda\cdot\vec u\big)\vec u}{c(u^o+c)}\Big]\frac{a^o}{c}=-\frac{1}{c}(P_o \vec a-\vec P a_o)\equiv-\vec K \frac{|\vec\lambda|}{c}\,.
\end{align*}

In particular, we observe the actual constraint defining the mass-shell: 
\[
\big(i_{\tilde X^R_{a^o}}\Theta\big)^2-\big(i_{(\tilde X^R_{\vec a}+\vec \lambda\Xi)}\Theta\big)^2=P_o^2-\vec P^2=-\vec\lambda^2
\,.
\]

It is convenient to redefine slightly the Noether invariants in order to have a more canonical set of coordinates and momenta. Let us reorganize the generators in this way: 
\[
\big\lbrace \vec\lambda\cdot \tilde X_{\vec a}\,,\vec\lambda\cdot \tilde X_{\vec \epsilon}\,,\vec\lambda\wedge \tilde X_{\vec u}\,; \;\;
	\tilde X_{a^o}\,,\vec\lambda\wedge \tilde X_{\vec a}\,;\;\;\vec\lambda\wedge \tilde X_{\vec \epsilon}\,,\vec\lambda\cdot \tilde X_{\vec u}\,\big\rbrace\,.
\]

The left version of the first set generate the characteristic subgroup, which is constituted by the proper evolution generator $ \vec\lambda\cdot \tilde X_{\vec a}$ and the ``little group'' of the representation, with structure of $SO(2,1)$. The right version of the second set give rise, through the corresponding Noether invariants, to momenta, whereas the third set produce the conjugate coordinates. We then rewrite the Noether invariants as follows: 
\begin{itemize}
	\item $\mathcal C_{\Theta}$ (right-version of the characteristic subalgebra of $\Theta$): 
\begin{align*}
	\frac{\vec\lambda}{|\vec\lambda|}\cdot \Big(i_{(\tilde X^R_{\vec a}+\vec \lambda\Xi)}\Theta \Big) &= 
	\Big[R(\vec \epsilon\,)\vec \lambda + \frac{\big(R(\vec \epsilon\,)\vec \lambda\cdot\vec u\big)\vec u}{c(u^o+c)}\Big]\cdot \frac{\vec\lambda}{|\vec\lambda|} \equiv H \equiv P_{\parallel}\,, 
	\quad \hbox{or} \quad P_\parallel=  \frac{\vec\lambda}{|\vec\lambda|} \cdot \vec P\,,
\\
	\frac{\vec\lambda}{|\vec\lambda|}\cdot i_{\tilde X^R_{\vec \epsilon}}\Theta &=
	\Big\lbrace\vec a \wedge \Big[R(\vec \epsilon\,) \vec \lambda + \frac{\big(R(\vec \epsilon\,)\vec \lambda\cdot\vec u\big)\vec u}{c(u^o+c)}\Big]\Big\rbrace\cdot \frac{\vec\lambda}{|\vec\lambda|} \equiv J_{\parallel}\,,
	\quad \hbox{or} \quad J_\parallel=  \frac{\vec\lambda}{|\vec\lambda|} \cdot \vec J\,,
\\
	-\frac{\vec\lambda}{|\vec\lambda|}\wedge i_{\tilde X^R_{\vec u}}\Theta &= 
	\frac{\vec\lambda}{|\vec\lambda|}\wedge
	\Big\lbrace\frac{1}{c^2}\big(R(\vec \epsilon\,)\vec \lambda\cdot\vec u\big)\vec a - \Big[R(\vec \epsilon\,) \vec \lambda + \frac{\big(R(\vec \epsilon\,)\vec \lambda\cdot\vec u\big)\vec u}{c(u^o+c)}\Big]\frac{a^o}{c}\Big\rbrace \equiv 
	\vec K_{\perp} \frac{|\vec\lambda|}{c}\,,
	\\
	&\quad \hbox{or} \quad \vec K_\perp=  \frac{\vec\lambda}{|\vec\lambda|} \wedge \vec K\,.
\end{align*}

\item Coordinates: 
\begin{align*}
	\frac{\vec\lambda}{|\vec\lambda|}\wedge i_{\tilde X^R_{\vec \epsilon}}\Theta &=
	\frac{\vec\lambda}{|\vec\lambda|}\Big\lbrace\vec a \wedge \Big[R(\vec \epsilon\,) \vec \lambda + \frac{\big(R(\vec \epsilon\,)\vec \lambda\cdot\vec u\big)\vec u}{c(u^o+c)}\Big]\Big\rbrace \equiv \vec J_{\perp}\,,
	\quad \hbox{or} \quad \vec J_\perp=  \frac{\vec\lambda}{|\vec\lambda|} \wedge \vec J\,,
\\
	-\frac{\vec\lambda}{|\vec\lambda|}\cdot i_{\tilde X^R_{\vec u}}\Theta &= 
	\frac{\vec\lambda}{|\vec\lambda|}\cdot
	\Big\lbrace\frac{1}{c^2}\big(R(\vec \epsilon\,)\vec \lambda\cdot\vec u\big)\vec a - \Big[R(\vec \epsilon\,) \vec \lambda + \frac{\big(R(\vec \epsilon\,)\vec \lambda\cdot\vec u\big)\vec u}{c(u^o+c)}\Big]\frac{a^o}{c}\Big\rbrace \equiv 
	 K_{\parallel} \frac{|\vec\lambda|}{c}\,,
	 \\
	 \quad &\hbox{or} \quad K_\parallel=  \frac{\vec\lambda}{|\vec\lambda|} \cdot \vec K\,.
\end{align*}

\item Momenta:
\begin{align*}
	- i_{\tilde X^R_{a^o}}\Theta &= R(\vec \epsilon\,)\vec \lambda \cdot \frac{\vec u}{c}\equiv P_o\,,\\
	\frac{\vec\lambda}{|\vec\lambda|}\wedge \Big(i_{(\tilde X^R_{\vec a}+\vec \lambda\Xi)}\Theta \Big) &= 
	\frac{\vec\lambda}{|\vec\lambda|}\wedge	\Big[R(\vec \epsilon\,)\vec \lambda + \frac{\big(R(\vec \epsilon\,)\vec \lambda\cdot\vec u\big)\vec u}{c(u^o+c)}\Big]  \equiv \vec P_{\perp}\,, 
	\quad \hbox{or} \quad \vec P_\perp=  \frac{\vec\lambda}{|\vec\lambda|} \wedge \vec P\,.
\end{align*}
\end{itemize}

Those invariants in $\mathcal C_{\Theta}$ can be rewritten in terms of the basic ones. In fact, 
\[
H=\pm \sqrt{\lambda^2+P_o^2-\vec P_\perp^2}\,,\quad \vec K_\perp=\frac{1}{H}\big(K_\parallel \vec P_\perp-P_o \vec\lambda\wedge\vec J_\perp\big)\,,\quad J_\parallel=\frac{1}{H}\big(\vec P_\perp\cdot\vec J_\perp \big )\,, 
\]
where both signs of the square root must be taken into account in order to cover the whole one-sheeted hyperboloid (note that, for tardions, only one sign, either $+$ or $-$, is required to cover each one of the orbits of the two-sheeted hyperboloid). 

It should be highly remarked the non-trivial fact that the Noether invariants parametrizing the momentum space, that is, $P_o$, $\vec P_\perp$, are functions not only of the boost parameters of the Poincar\'e group, but also of rotations. In particular, $\vec P_\perp$ depends primarily on the rotation parameters. Similarly, the position invariants depend on the rotation parameters at the same level than translation ones.
The quantization in momentum space then proceeds by taking a first-order Polarization subalgebra constituted by $\mathcal C_\Theta$ along with the left version of the generators of momenta: 
\[
\mathbb P = \langle \vec \lambda \cdot \tilde X^L_{\vec a}\,,\vec\lambda\cdot\tilde X^L_{\vec \epsilon}\,,\vec \lambda\wedge\tilde X^L_{\vec u}\,;\tilde X^L_{a^o}\,,\vec\lambda\wedge\tilde X^L_{\vec a}\rangle\,.
\] 

The polarization condition on complex $U(1)$-functions on $\tilde{\mathcal P}$, $\Psi\equiv \zeta \psi(g)$, leads (after a non-trivial computation) to the wave functions
\[
\Psi=\zeta e^{-i R(\vec \epsilon\,)\vec \lambda \cdot \frac{\vec u}{c} a^o} e^{i  \big[(R(\vec \epsilon\,)-\mathbb I)\vec \lambda + \frac{(R(\vec \epsilon\,)\vec \lambda\cdot\vec u)\vec u}{c(u^o+c)}\big]\cdot\vec a}
\Phi \big(R(\vec \epsilon\,)\vec \lambda \cdot \frac{\vec u}{c},\,R(\vec \epsilon\,)\vec \lambda \wedge \frac{\vec\lambda}{|\vec\lambda|}+ \frac{(R(\vec \epsilon\,)\vec \lambda\cdot\vec u)\vec u\wedge\frac{\vec\lambda}{|\vec\lambda|}}{c(u^o+c)}\big)\,,
\]
that is to say, an arbitrary function $\Phi$ of $P_o$, $\vec P_\perp$ multiplied by specific functions containing coordinates $a_o$, $\vec a$. 

Let us indicate briefly the mathematical steps leading to this result. Since the polarization contains $\vec\lambda\cdot \tilde X^L_{\vec a}$ (in $\mathcal C_\Theta$) as well as $\tilde X^L_{a^o}$ and $\vec\lambda\wedge \tilde X^L_{\vec a}$, we may impose to $\Psi$ the conditions $\tilde X^L_{a^o}\Psi=0$, $\tilde X^L_{\vec a} \Psi=0$. This suggests that $\Psi$ must be of the form
\[
\Psi= \zeta e^{-i(\gamma_o a^o+ \vec \gamma \cdot \vec a)} \Phi(``\vec \lambda\cdot \vec u\, \hbox{''}\,,\, ``\vec \lambda\wedge \vec \epsilon\,\, \hbox{''})\,,
\]
where $\gamma_o$, $\vec \gamma$ can depend on $\vec u$ and $\vec \epsilon$ although in a fixed form, and by $``\vec \lambda\cdot \vec u\, \hbox{''}$ and  $``\vec \lambda\wedge \vec \epsilon\,\, \hbox{''}$ we mean that $\Phi$ depends on some arguments determined by the specific form of the polarization condition.

In order to give a glimpse of the calculations involved, we enter in some more detail here. By imposing $\tilde X^L_{a^o}\Psi=0$, $\tilde X^L_{\vec a} \Psi=0$ (or, easier, $R^{-1}(\vec \epsilon\,)\tilde X^L_{\vec a} \Psi=0$), we arrive at 
\begin{align*}
	\gamma^o &= \frac{\vec u}{u^o}\cdot(\vec \lambda-\vec \gamma)\,,\quad 
	\vec \gamma +\frac{\vec u \cdot(\vec \lambda-\vec \gamma)}{u^o(u^o+c)}\vec u-
	(\mathbb I-R(\vec \epsilon\,))\vec \lambda =0
\\
\Rightarrow
\gamma^o&= R(\vec \epsilon\,)\vec \lambda \cdot \frac{\vec u}{c} \,,\quad
\vec \gamma = (\mathbb I-R(\vec \epsilon\,))\vec \lambda - \frac{(R(\vec \epsilon\,)\vec \lambda\cdot\vec u)\vec u}{c(u^o+c)}\,.
\end{align*}

In the same way, the polarization condition $\vec\lambda\cdot \tilde X^L_{\vec \epsilon}\Psi=0$ implies 
\[
\vec \lambda \cdot X^{L\,SU(2)}_{\vec \epsilon} \big[(-i R(\vec \epsilon\,)\vec \lambda \cdot \vec u)a^o+iu^o R(\vec \epsilon\,)\vec \lambda \cdot \vec a \big]\psi + e^{-i R(\vec \epsilon\,)\vec \lambda\cdot \frac{\vec u}{c}a^o}e^{i  \big[(R(\vec \epsilon\,)-\mathbb I)\vec \lambda + \frac{(R(\vec \epsilon\,)\vec \lambda\cdot\vec u)\vec u}{c(u^o+c)}\big]\cdot\vec a} \vec \lambda \cdot X^{L\,SU(2)}_{\vec \epsilon} \Phi=0
\]
\[
\Rightarrow \vec \lambda \cdot X^{L\,SU(2)}_{\vec \epsilon} \Phi=0\, ,
\]
since the first term is zero as a consequence of being zero  $\vec \lambda \cdot X^{L\,SU(2)}_{\vec \epsilon} R(\vec \epsilon\,)\vec \lambda$. The remaining condition, $\vec \lambda \cdot X^{L\,SU(2)}_{\vec \epsilon} \Phi=0$, says that $\Phi = \Phi (\hbox{``}\vec \lambda\cdot\vec u\,\hbox{''}, R(\vec \epsilon\,)\vec \lambda)$. 

Finally, the condition
\[
\vec \lambda\wedge\tilde X^L_{\vec u}\psi=0 \quad \Rightarrow \quad 
\Phi = \Phi \Big(R(\vec \epsilon\,)\vec \lambda\cdot \frac{\vec u}{c}, 
 R(\vec \epsilon\,)\vec \lambda \wedge\frac{\vec \lambda}{|\vec \lambda|}+\frac{(R(\vec \epsilon\,)\vec \lambda\cdot\vec u )\vec u\wedge\frac{\vec \lambda}{|\vec \lambda|}}{c(u^o+c)}\Big). 
\]

The action of the right generators on $\Psi$ can be projected to an action on just the arbitrary functions $\Phi$, giving rise to the actual physical quantum operators in momentum space representation, that is: 
\begin{align*}
	i\tilde X^R_{a^o} \Phi &= \big (R(\vec \epsilon\,)\vec\lambda\cdot \vec u \big )\Phi
\\
	i \tilde X^R_{\vec a} \Phi  &= \Big( -(R(\vec \epsilon\,)-\mathbb I)\vec \lambda - \frac{(R(\vec \epsilon\,)\vec \lambda\cdot\vec u)\vec u}{c(u^o+c)}\Big)\Phi
\\
	i \tilde X^R_{\vec u}  \Phi &  = -i \frac{P_o}{|\vec \lambda|}\frac{\partial \Phi}{\partial \vec P}-i\frac{\vec P}{|\vec \lambda|}\frac{\partial \Phi}{\partial P_o} 
\\
	i \tilde X^R_{\vec \epsilon}\Phi  &=  -i \vec P \wedge  \frac{\partial \Phi}{\partial \vec P}\,,
\end{align*}
and the natural basic operators on the ``momentum space'':  
\begin{itemize}
	\item Coordinate operators: 
	\begin{align*}
		 i \tilde X^R_{\vec \epsilon_\perp}&= P_\parallel \frac{\vec \lambda}{|\vec \lambda|}\wedge \frac{\!\!\partial}{\partial \vec P_\perp} \equiv \hat{\vec{J}}_\perp \,,
		 \\
		 \frac{\vec \lambda}{|\vec \lambda|}\cdot i \tilde X^R_{\vec u}&= 
		 -i \frac{P_o}{|\vec \lambda|}\vec \lambda \cdot\frac{\!\!\partial}{\partial \vec P}-i\frac{P_\parallel}{|\vec \lambda|}\frac{\!\!\partial}{\partial P_o} = -i\frac{P_\parallel}{|\vec \lambda|}\frac{\!\!\partial}{\partial P_o} \equiv \hat K_\parallel\,.
	\end{align*}
	\item Momentum operators: 
	\begin{align*}
	-i\tilde X^R_{a^o} &= R(\vec \epsilon\,)\vec\lambda\cdot \vec u \equiv \hat P_o\,,
\\
	\frac{\vec \lambda}{|\vec \lambda|}\wedge i (\tilde X^R_{\vec a} + c \vec \lambda \Xi)  &= \frac{\vec \lambda}{|\vec \lambda|}\wedge \Big( -(R(\vec \epsilon\,)-\mathbb I)\vec \lambda - \frac{(R(\vec \epsilon\,)\vec \lambda\cdot\vec u)\vec u}{c(u^o+c)}\Big) \equiv \hat{\vec P}_\perp\,,
\end{align*}
\end{itemize}
along with the derived ones (those on the characteristic subgroup): 
\[
\hat H = \sqrt{\lambda^2+ \hat P_o^2-\hat{\vec P}_\perp^2}\,,\quad 
\hat{\vec K}_\perp=\frac{1}{H}\big(\hat K_\parallel \hat{\vec P}_\perp-\hat P_o \vec\lambda\wedge\hat{\vec J}_\perp \big)\,, \quad
\hat J_\parallel=\frac{1}{H} \big(\hat{\vec P}_\perp \cdot \hat{\vec J}_\perp\big)\,.
\]

\section{Configuration space representation}

In this Section, we just indicate briefly the way the more conventional and (supposedly) better known configuration space representation is carried out under the present group approach, without entering into explicit computations. This representation is actually more intricate than it might appear at first sight due to the peculiar scalar product involved.

As in the standard two-sheeted hyperboloid case ($m^2>0$), the representation in configuration space does not come from a first-order polarization. In fact, in order to achieve a wave function not depending on other variables than coordinates, we should be provided by polarization conditions of the form: 
\[
 \langle \vec \lambda \cdot \tilde X^L_{\vec a}\,,\vec \lambda \cdot\tilde X^L_{\vec \epsilon}\,,\vec \lambda \wedge\tilde X^L_{\vec u}\,;
 \vec \lambda \cdot \tilde X^L_{\vec u}\,,\vec \lambda \wedge \tilde X^L_{\vec \epsilon}\rangle
 \equiv
  \langle \mathcal C_\Theta\,;
 \vec \lambda\cdot \tilde X^L_{\vec u}\,,\vec \lambda \wedge \tilde X^L_{\vec \epsilon}\rangle
 \,, 
\]
so that the wave functions only could depend on $(a_o,\vec a_\perp)$. Unfortunately, such a set of generators do not close a subalgebra. The simplest solution consists in resorting to a higher-order polarization preserving the maximum of the desired generators; in particular, $\vec \lambda \cdot \tilde X^L_{\vec u}$ and  $\vec \lambda \wedge \tilde X^L_{\vec \epsilon}$. This higher-order polarization is 
\[
\mathbb P^{HO}\equiv
 \langle \tilde X^{L\,HO}_{a_\parallel}\,,\vec \lambda \cdot\tilde X^L_{\vec \epsilon}\,,\vec \lambda \wedge\tilde X^L_{\vec u}\,;
 \vec \lambda\cdot \tilde X^L_{\vec u}\,,\vec \lambda \wedge \tilde X^L_{\vec \epsilon}\rangle\,,
\]
where we have replaced $\vec \lambda \cdot \tilde X^L_{\vec a}$ by 
\[
\tilde X^{L\,HO}_{a_\parallel}\equiv \tilde X^{L\,2}_{a^o}-\tilde X^{L\,2}_{\vec a_\perp}-\big(\tilde X^L_{a_\parallel}+\lambda\Xi\big)^2+\lambda^2\Xi^2\,.
\]
By imposing $\mathbb P^{HO}$ on $\Psi$ and restoring the ``rest mass'' on $\Psi$, that is, redefining $\Psi = e^{-i\vec \lambda\cdot\vec a}\Phi(a_o,\vec a_\perp,a_\parallel)$, $\Phi$ evolves on $a_\parallel$ according to the Klein-Gordon equation
\[
\Big(\frac{\partial^2}{\partial a_o^2}-\frac{\partial^2}{\partial {\vec a_\perp}}-\frac{\partial^2}{\partial a_\parallel^{\,2}}-\lambda^2\Big)\Phi=0\,.
\]
It should be remarked that, even though $a^o$ and $a_\parallel$ appear in the equation in the same footing, the evolution takes place really in $a_\parallel$ as can be realized once the Galilean limit is taken, where the derivative with respect to $a_o$ actually disappear (see \cite{galitaquis}).

The action of the right-invariant vector fields on $\Psi$ is easily reduced to the corresponding action on mere configuration space. That is
\begin{align*}
	i\tilde X^R_{a^o} \Psi &= i e^{-i \vec \lambda a \cdot \vec a}  \frac{\partial \Phi}{\partial a_o}
\\
	i \tilde X^R_{\vec a} \Psi  &= i e^{-i \vec \lambda a \cdot \vec a}  \frac{\partial \Phi}{\partial \vec a}+ \vec \lambda e^{-i \vec \lambda a \cdot \vec a}  \Phi \qquad \hbox{or} \qquad \Big(i \tilde X^R_{\vec a} + i \vec \lambda \Xi \Big)\Psi =  i e^{-i \vec \lambda a \cdot \vec a}  \frac{\partial \Phi}{\partial \vec a}
\\
	i \tilde X^R_{\vec u}  \Psi &  = i e^{-i \vec \lambda a \cdot \vec a} \Big(\frac{a^o}{c}  \frac{\!\!\partial}{\partial \vec a}+\frac{\vec a}{c}  \frac{\!\!\partial }{\partial  a^o}\Big)\Phi
\\
	i \tilde X^R_{\vec \epsilon}\Psi  &=  i e^{-i \vec \lambda a \cdot \vec a} \vec a \wedge  \frac{\partial \Phi}{\partial \vec a}\,,
\end{align*}
and the natural operators on $\Phi$ become: 
\begin{align*}
	\hat P_o \Phi &= i  \frac{\partial \Phi}{\partial a^o}
	\\
	\hat{\vec P}_\perp \Phi &= i \vec \lambda \wedge  \frac{\partial \Phi}{\partial \vec a}
	\\
	\hat{\vec J}_\perp \Phi &= i \frac{\vec \lambda}{|\vec \lambda|}\wedge \Big(\vec a \wedge  \frac{\partial \Phi}{\partial \vec a}\Big)
	\\
	\hat K_\parallel \Phi &=i \frac{a^o}{c}\vec \lambda\cdot  \frac{\partial \Phi}{\partial \vec a}+i\frac{\vec \lambda\cdot\vec a}{c} \frac{\partial \Phi}{\partial a_o} \equiv i\frac{a^o}{c}\frac{\partial \Phi}{\partial a_\parallel}+i\frac{a_\parallel}{c} \frac{\partial \Phi}{\partial a_o}
\end{align*}

Again, the evolution parameter turns out to be $a_\parallel$ and now, properly, the Cauchy surface on which $\Phi$ takes initial values proves to be that spanned by $(a_o,\vec a_\perp)$ or other (time-like) Lorentz-equivalent. 

Let us remind the reader that the configuration space representation already counts with a proper scalar product which makes plausible the physical consideration of tachyons \cite{PRD}.

\section{Conclusions and outlook}

Traditionally, tachyons are treated just as particles with imaginary mass, thus leading to many inconsistencies which ended up with their exclusion from Particle Physics. The present analysis demonstrates that this is not properly the situation with elementary systems described by representations of the Poincar\'e group associated with the one-sheeted hyperboloid. The difference with ordinary particles is deeper and relies basically in the fact that the evolution parameter is not the ordinary time, but one spatial direction $(\vec x \parallel \vec \lambda)$. This means that (in configuration space) ordinary time $x^o\equiv a^o$ belongs to the Cauchy surface of initial conditions for the system (either classical or quantum). 

Under this situation, the traditional analysis of causality and unitarity must be soundly revised. In particular, a novel dynamics arises on which a tachyonic system is allowed to span the spatial size at constant ordinary time. Thinking of simple, cosmological evolution, this means that an inflationary scenario is quite natural and actually physical without a loss of unitarity nor causality. Although, on the other hand, a universe with tachyons would not need a specific inflation program due to the lack of horizon. In a similar manner, taking tachyonic matter to Cosmology opens the door to gravitational scenarios generated by ``negative energy'' conditions, such as non-transversable worm holes, etc.

As far as the tachyonic quantum field theory is concerned, there seems not to find structural obstructions provided that we maintain the assumption that tachyons evolve in a spatial direction. The solutions of the Klein-Gordon equation, $\mathcal H$, should be characterized by the initial values of (integrable) functions on the Cauchy surface and derivatives of functions with respect to the evolution parameter $\vec \lambda \cdot \vec x$ on the same surface. 

A group approach to QFT of tachyons would proceed by constructing an infinite-dimensional group containing, apart from the kinematical symmetry group $\mathcal P$, a Heisenberg-Weyl group, which is a central extension by $U(1)$ of $\mathcal H \otimes \mathcal H^*$ (where $\mathcal H^*$ is the dual space of $\mathcal H$), as was made in the bradyonic case (see \cite{CMP,IJMP,EurPhys}).

In recent papers \cite{Dragan1,Dragan2} a duplication of the Hilbert space of the free tachyons $\mathcal F \otimes \mathcal F^*$ was invoked to solve the invariance of the whole Hilbert space under boosts, which may change the sign in the exponential in $i x_o$, changing positive energy solutions into negative energy ones, or input into output states. This trouble is again a consequence of keeping the traditional (bradyonic) Cauchy surface. 

A tentative construction of a QFT of tachyons according to the present Group Approach to Quantization scheme will be published elsewhere.

\section*{Appendix}

Group Approach to Quantization starts by looking for central extensions by $U(1)$, $\tilde G$, of the classical symmetry group $G$ of a given physical system. They are characterized by a cocycle \cite{23} $\Delta\equiv e^{i \xi}: G\times G\rightarrow U(1)$. The function $\xi$ defines a bilinear form $\Sigma: \mathcal G \times \mathcal G \rightarrow \mathbb R$ on the Lie algebra $\mathcal G$ whose kernel, $\mathcal C \equiv Ker \Sigma$, points out the generalized equations of motion, so that, the algebraic quotient $\tilde G / \mathcal C$ proves to be a Quantum Manifold in the sense of Geometric Quantization \cite{GQ1,GQ2,GQ3}. 

The actual quantization takes place when the right-invariant generators $\tilde X^R \in \tilde{\mathcal X}^R\equiv \tilde{\mathcal G}$ act as operators on complex $U(1)$-functions on $\tilde G$, previously restricted to depend arbitrarily on half the variables of $\tilde G/(\mathcal C \cup U(1))$. This restriction is accomplished by means of a Polarization, that is, a maximal subalgebra of $\tilde{\mathcal X}^L\equiv \mathcal G$ containing $\mathcal C$. This restriction is compatible with the action of the operators since $\tilde{\mathcal X}^L$ and $\tilde{\mathcal X}^R$ commute for any Lie group. The existence of a Polarization containing the whole $\mathcal C$ (ful Polarization) is not guaranteed but, in that case, we can look for a higher-order Polarization in the left enveloping algebra. Higher-order Polarizations are also used even if a first-order Polarization does exist but we desire a different realization of the quantization. Usually, the configuration space representation requires a higher-order Polarization. The advantage of a first-order Polarization lies in the existence of an invariant measure for the scalar product \cite{polas}. 

A preponderant role is played by the $U(1)$-component of the left invariant canonical 1-form $\Theta\equiv \theta^{L\,U(1)}$, which generalizes the Poincar\'e-Cartan 1-form in Cartan Mechanics. \cite{Cartan}. In fact, $\mathcal C$, realized by means of left-invariant vector fields, $\mathcal C_\Theta$, constitutes the set of generalized classical equations of motion, on which the Noether invariants, $i_{\tilde X^R_{()}}\Theta$, are constant.

\bigskip

V.A. thank the Spanish Ministerio de Ciencia e Innovaci\'on (MICINN) for financial support (PID2022-116567GB-C22). J.G. and FF.LR. acknowledge financial support from the Spanish MICINN (PID2022-138144NB-100).  V.A. also acknowledges financial support from the State Agency for Research of the Spanish MCIU through the `Center of Excellence Severo Ochoa' award for the Instituto de Astrof\'\i sica de Andaluc\'\i a (SEV-2017-0709).


%
%

%



\end{document}